\documentclass[a4paper,fleqn,usenatbib]{mnras}

\usepackage[T1]{fontenc}
\usepackage{ae,aecompl}

\usepackage{graphicx}	
\usepackage{amsmath}	
\usepackage{amssymb}	

\title[Raman O~VI in AG~Dra]{Stellar Wind Accretion and Raman Scattered O~VI Features 
in the Symbiotic Star AG~Draconis}

\author[Y.-M. Lee et al.]{
Young-Min Lee,$^{1}$\thanks{E-mail: ymlee9211@gmail.com}
Hee-Won Lee,$^{1}$\thanks{E-mail: hwlee@sejong.ac.kr}
Ho-Gyu Lee$^{2}$
and Rodolfo Angeloni$^{3,4}$
\\
$^{1}$ Department of Physics and Astronomy, Sejong University, Seoul, Korea\\
$^{2}$ Korea Astronomy and Space Science Institute, Daejeon, Korea\\
$^{3}$ Instituto de Investigaci\'on Multidisciplinar en Ciencia y Tecnolog\'ia, Universidad de La Serena, Av. Ra\'ul Bitr\'an 1305, La Serena, 
Chile\\ 
$^{4}$ Departamento de F\'isica y Astronom\'ia, Universidad de La Serena, Av. Cisternas 1200 Norte, La Serena, Chile 
}

\date{Accepted 2019 May 14. Received 2019 May 13; in original form 2019 March 05}

\pubyear{2019}

\begin{document}
\label{firstpage}
\pagerange{\pageref{firstpage}--\pageref{lastpage}}
\maketitle

\begin{abstract}
We present high resolution spectroscopy of the yellow symbiotic star AG Draconis with 
ESPaDOnS at the {\it Canada-France-Hawaii Telescope}. 
Our analysis is focused on the profiles of Raman scattered \ion{O}{VI} features centered at 6825 \AA\ and 7082 \AA,
which are formed through Raman scattering of \ion{O}{VI}$\lambda\lambda$1032 and 1038 
with atomic hydrogen.  These features are found to exhibit double component profiles with conspicuously enhanced red parts. 
Assuming that the \ion{O}{vi} emission region constitutes a part of the accretion flow around the white dwarf, 
Monte Carlo simulations for \ion{O}{VI} line radiative transfer are performed to find that
the overall profiles are well fit with the accretion flow
azimuthally asymmetric with more matter on the entering side than on the opposite side.
As the mass loss rate of the giant component is increased, we find that the flux ratio $F(6825)/F(7082)$  of Raman 6825 and 7082 features decreases
and that our observational data are consistent with a mass loss rate $\dot M\sim 2 \times 10^{-7}
{\rm\ M_{\odot}\ yr^{-1}}$. We also find that additional bipolar components moving away with a speed 
$\sim 70{\rm\ km\ s^{-1}}$ provide considerably improved fit 
to the red wing parts of Raman features. The possibility that the two Raman profiles differ is briefly discussed in relation to
the local variation of the \ion{O}{VI} doublet flux ratio. 
\end{abstract}


\begin{keywords}
radiative transfer -- scattering --  (stars) binaries: symbiotic --  accretion disks -- stars-individual(AG Dra)
\end{keywords}



\section{Introduction}

A symbiotic star is a wide binary system consisting of a white dwarf and a late type giant undergoing
heavy mass loss \citep[e.g.,][]{kenyon86}. The cool giant component loses its material in the form of a slow stellar wind, 
some fraction of which is gravitationally captured by the white dwarf component giving rise to a variety
of activities including X-ray emission, outbursts and prominent emission lines.
The mass loss rate of the cool component in a typical symbiotic star has been proposed to be in the range 
$10^{-8}- 10^{-6}\  \rm{M_{\odot}\ yr^{-1}}$ \citep[e.g.,][]{dupree86}. 


AG~Draconis is known to be a yellow symbiotic star having an early K type giant 
as mass donor \citep[e.g.][]{leedjarv16}. 
The orbital period is known to be 550 days \citep[e.g.][]{fekel00}. 
The light curve of AG~Dra shows that it underwent major outbursts in intervals of 12-15 years
with more frequent minor outbursts with a time scale of $\sim 1$ year \citep{hric14}. 
\cite{sion12} proposed the effective temperature
of the hot component $T_{\rm eff}=80,000{\rm\ K}$ based on their analysis of spectra obtained
with the {\it Far Ultraviolet Spectroscopic Explorer (FUSE)}. 
\cite{smith96} investigated the chemical abundances of AG~Dra to show that it is a metal
poor star belonging to a halo population with overabundant  heavy $s$-process
elements. 

The presence of an accretion disk is well established in cataclysmic variables, where the primary white dwarf
is accreting material from the red dwarf companion filling its Roche lobe. However, in the case of
symbiotic stars, there is lack of direct evidence indicating the presence of an accretion disk. 
\cite{leibowitz85} suggested the presence of an accretion disk in AG~Dra
based on the flat continuum. \cite{robinson94} proposed that an accretion disk is responsible for the double-peak profiles of H$\alpha$
observed in a number of symbiotic stars including AG~Dra. However, they also noted the mismatch of H$\alpha$ profiles
with the accretion disk model suggesting an alternative model, where H$\alpha$ is formed in the Str{\"o}mgren sphere
and suffers self-absorption in the outlying neutral region of the red giant wind.

A number of symbiotic stars show evidence of bipolar outflows that may be closely related to the
formation of an accretion disk around the hot white dwarf component.
The radio interferometric observations using the {\it Multi-Element Linked Interferometer Network,  (MERLIN)} 
by \cite{mikolajewska02} revealed a bipolar structure 
that is well aligned with the binary plane of AG~Dra. A similar result was  presented by
\cite{torbett87}.
Despite the indication of a bipolar structure in AG~Dra, observational verification
of the presence of an accretion disk in this system still remains a difficult issue.  

The formation of an accretion disk can be addressed through hydrodynamic calculations.
\cite{mastrodemos98} carried 
out Smoothed-Particle Hydrodynamics (SPH) calculations to propose that  
a stable accretion disk may form around a white dwarf that accretes material from a giant companion. 
The two dimensional hydrodynamical simulations performed by \cite{devalborro09} revealed  
 that stable accretion disks can 
form with mass loss ranging $10^{-9}-10^{-6}{\rm\ M_\odot\ yr^{-1}}$.
In particular, they noted that the accretion flow is characterized by eccentric streamlines and Keplerian velocity profiles.

Raman scattered \ion{O}{vi} features at 6825 \AA\ and 7082 \AA\ that have been found only in bona fide symbiotic stars 
including AG~Dra are unique tools to probe the accretion flow in symbiotic stars.
These mysterious spectral features were
identified by \cite{schmid89}, who proposed that they are formed through
Raman scattering of \ion{O}{VI}$\lambda\lambda$1032 and 1038 by atomic hydrogen. 
When a far UV \ion{O}{VI} line photon is incident on a hydrogen atom in the ground 
state, the hydrogen atom is excited to one of infinitely many $p$ states.
There are two channels of de-excitation available for the excited hydrogen atom. One 
is de-excitation into the ground $1s$ state, which corresponds to Rayleigh scattering. 
The other channel is Raman scattering, where the final de-excitation is made into the 
excited $2s$ state with an emission of an optical photon redward of H$\alpha$.

Many Raman scattered \ion{O}{VI} features show double or triple 
peak profiles. 
\cite{lee07} undertook profile analyses of the two symbiotic
stars V1016~Cygni and HM~Sagittae to show that the double peak profiles in these 
objects are well fit by adopting a Keplerian accretion disk around the hot white 
dwarf component \citep[see also][]{heo15, heo16}.  The fitted profiles are consistent 
with the \ion{O}{VI} emission region in Keplerian motion with $\sim
50{\rm\ km\ s^{-1}}$ implying the physical dimension of the accretion disk  
$\le 1{\rm\ au}$. In addition, they attributed the red enhanced asymmetric profiles
to higher concentration of emitting material on the entering side of the accretion flow than on the opposite side.

In this paper, we present our high resolution spectrum of AG~Dra showing Raman \ion{O}{VI} features.
In view of their great usefulness as a probe of 
the \ion{O}{VI} emission region, we perform Monte Carlo profile analyses to propose 
that the Raman \ion{O}{VI} profiles are consistent with  the presence of an accretion disk in this binary system.

\section{Spectroscopic Observation of AG~Dra}

\subsection{{\it CFHT} Spectrum of AG~Dra}
\begin{figure*}
\centering
\includegraphics[scale=.45]{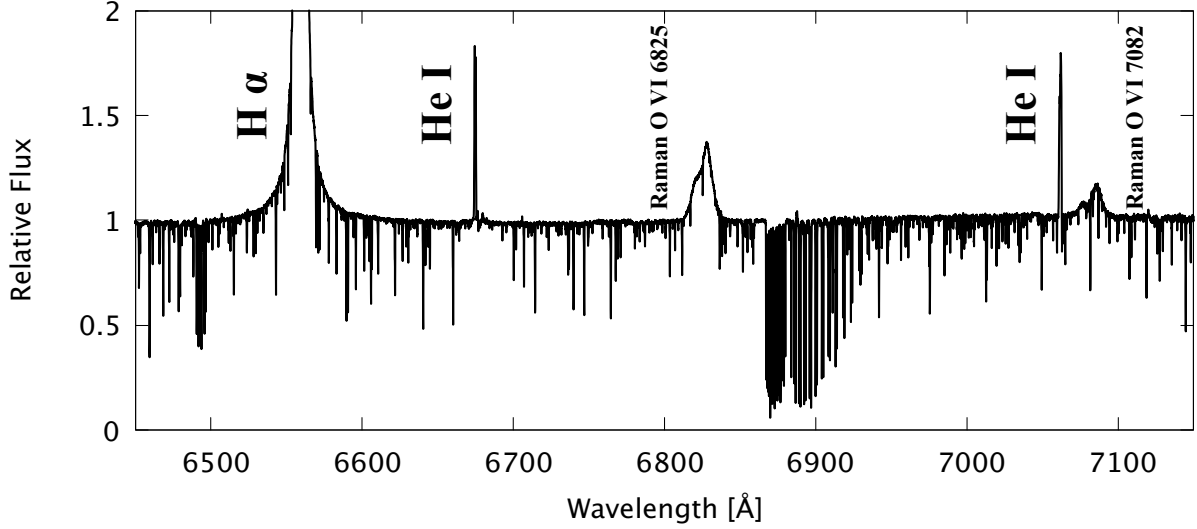}
\caption{High resolution spectrum of  AG~Dra obtained with the echelle spectrograph ESPaDOnS and {\it CFHT}
showing Raman scattered \ion{O}{VI}$\lambda\lambda$ 1032 and 1038 at
around 6825 \AA\ and 7082 \AA\ as well as H$\alpha$, \ion{He}{I}$\lambda$6678, and \ion{He}{I}$\lambda$7062.
It is also notable that O$_2$ telluric B band absorption is prominently seen redward of
6867 \AA.
}
\label{cfht}
\end{figure*}

We carried out high resolution spectroscopy of AG~Dra using the
{\it Echelle SpectroPolarimetric Device for the Observation of Stars} 
(ESPaDOnS) installed on the 3.6 m {\it Canada-France-Hawaii Telescope
(CFHT)} 
in queued service observing mode on  2014 September 6.  
The total integration time was 2,000 seconds and 
the spectral resolution is 81,000.
Data reduction was performed using the {\it CFHT}
reduction pipeline Upena, which is based on the package
Libre-ESpRIT developed by \cite{donati97}.
A standard Th-Ar lamp was used for wavelength calibration, 
according to which the pipeline wavelength should be shifted
by  $-0.18{\rm\ km\ s^{-1}}$ for heliocentric correction.

In Fig.~\ref{cfht}, we show the part of the spectrum in the wavelength range between 6460 \AA\ and 7140 \AA,
in which we find Raman scattered \ion{O}{vi} features as well as H$\alpha$, two \ion{He}{i} emission lines
at 6678 \AA\ and 7062 \AA\ and
many absorption features due to the giant atmosphere.
The vertical axis shows the relative flux, where the normalization is made in such a way that the continuum level is of unit value.

\begin{figure*}
\centering
\includegraphics[scale=0.48]{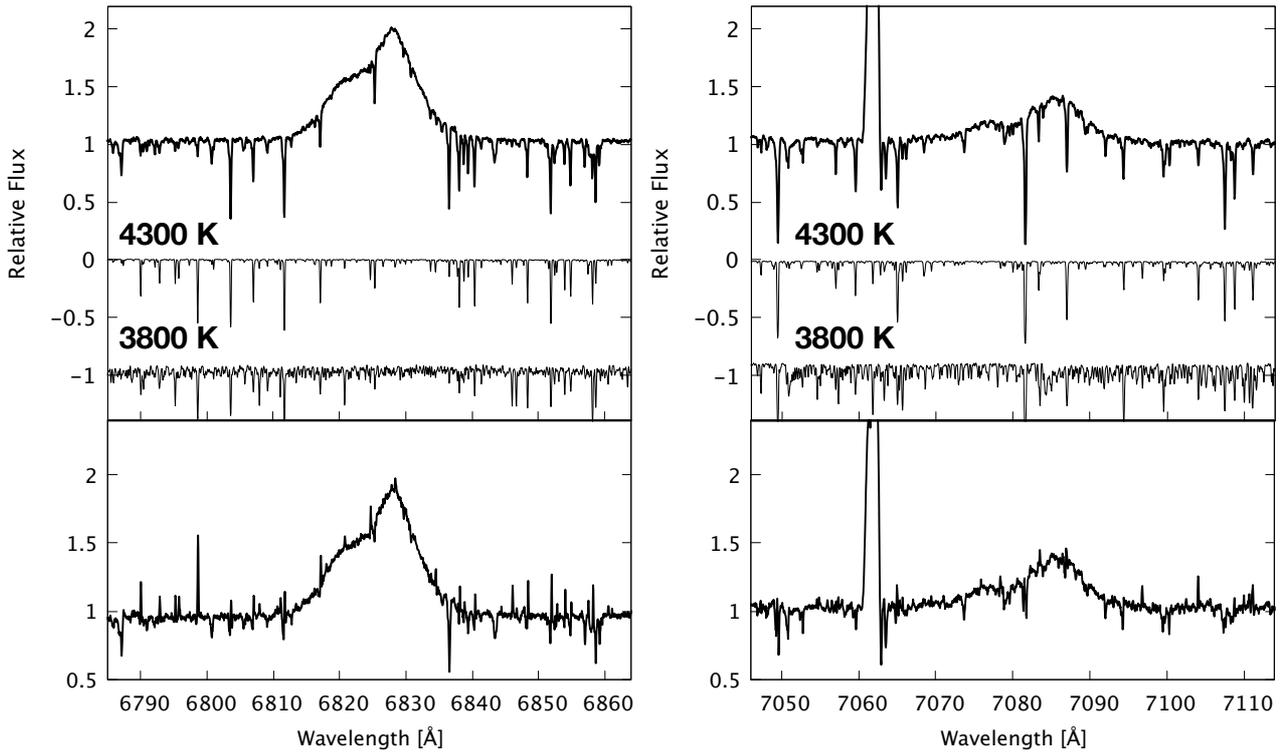}
\caption{{\it CFHT} spectra around Raman O VI 6825 and 7082 features of AG~Dra (top panels) 
and the BT-Settl model atmospheres of a giant star with $T_{\rm eff}=4,300$ and $3,800{\rm\ K}$.
The bottom panels show the spectra corrected for the BT-Settl model 
atmosphere with $T_{\rm eff}=4,300{\rm\ K}$.  
}
\label{btsettl}
\end{figure*}

The upper left and right panels of Fig.~\ref{btsettl} show the parts of the spectrum around 
Raman \ion{O}{VI} features centered at 6825 \AA\ and 7082 \AA, respectively. In these panels, the continuum is subtracted
and the flux is rescaled so that the red peak of the Raman 6825 is of unit strength.

 We note that the spectrum of AG~Dra is plagued with many absorption features attributed to the atmosphere of the giant component.
 In order to assess the effect of the stellar atmospheric absorption lines on the overall profiles of the Raman \ion{O}{VI} features, 
we invoke the BT-Settl model atmosphere \citep{allard14}.
The results are presented in Fig.~\ref{btsettl}. In this figure, the top spectra are the blow-up version around the Raman
features of the {\it CFHT} data shown in Fig.~1.
The second and third
spectra show the BT-Settl model atmospheres of a yellow giant with $T_{\rm eff}=4,300$ and $3800{\rm\ K}$, respectively.
As is found in the figure, the effect of molecular absorption features is more serious in the case of $T_{\rm eff}=3,800{\rm\ K}$
than $T_{\rm eff}=4,300{\rm\ K}$.  

In the bottom we show the resultant spectra obtained by subtracting the model atmosphere with
$T_{\rm eff}=4,300{\rm\ K}$ from the observational data. 
The Raman 6825 feature is quite strong and affected by relatively weak absorption features,
exhibiting an overall smooth profile.  In contrast, in the case of the Raman 7082 feature,
the subtraction is reasonably good except a number of absorption/emission features
in the Doppler factor region $-25{\rm\ km\ s^{-1}}\le \Delta V<0{\rm\ km\ s^{-1}}$, 
which can be attributed to imperfect removal of continuum.
The imperfection of subtraction makes it very difficult 
to infer the exact amount of the Raman 7082 flux that is absorbed.

\subsection{Profile Comparison}

Raman and Rayleigh scattering of a far UV photon with an atomic electron is described by time dependent
second order perturbation theory \citep[e.g.][]{schmid89,lee97}.
The cross sections for \ion{O}{VI}$\lambda\lambda$1032
and 1038 are $\sigma_{\rm Ram}(1032)=7.5\sigma_{\rm T}$ and $\sigma_{\rm Ram}(1038)=2.5\sigma_{\rm T}$, 
respectively, where $\sigma_{\rm T}=0.665
\times 10^{-24}{\rm\ cm^2}$ is the Thomson scattering cross section 
\citep[e.g.,][]{schmid89, nussbaumer89, ylee16}. The Rayleigh scattering cross sections are found to be
$\sigma_{\rm Ray}(1032)=34\sigma_{\rm T}$ and $\sigma_{\rm Ray}(1038)=6.6\sigma_{\rm T}$.

The energy conservation principle sets a relation between the wavelength $\lambda_i$ of an incident 
photon and that $\lambda_o$ of Raman scattered one given by
\begin{equation}
\lambda_i^{-1}=\lambda_o^{-1}+\lambda_\alpha^{-1},
\label{wvl}
\end{equation}
where $\lambda_\alpha$ is the wavelength of Ly$\alpha$. According to \cite{schmid89}, 
the fiducial atomic line centers in air for Raman scattered \ion{O}{VI}
are 6825.44 \AA\ and 7082.40 \AA. Differentiation of Eq.~(\ref{wvl}) yields
\begin{equation}
{\Delta\lambda_o\over\lambda_o}=\left({\lambda_o\over\lambda_i} \right){\Delta\lambda_i\over\lambda_i},
\end{equation}
from which it may be noted that the Raman \ion{O}{VI} features will be broadened 
by a factor $(\lambda_o/\lambda_i)\sim 6$ with respect to the incident far UV \ion{O}{VI}. This relation dictates 
that Raman \ion{O}{VI} profiles
reflect relative motion between \ion{O}{VI} and \ion{H}{i} and that they
 are almost independent of the observer's line of sight. 
In view of this property, no consideration of transforming
to the binary center of mass frame is made in this work.

In the spectrum shown in Fig.\ref{btsettl}, the Raman 6825 and 7082 features are found to lie in the ranges
6810\ \AA \ $<\lambda<$\ 6840 \AA \ and 7068\ \AA \ $<\lambda
<7095$\ \AA, respectively, which translates into the full width at zero intensity (FWZI) of $200{\rm\ km\ s^{-1}}$ 
with respect to the neutral scattering region around the giant. If we interpret this velocity as the twice of the Keplerian
speed at the innermost \ion{O}{VI} emission region, then it implies that
\begin{equation}
r_{i} = 7\times 10^{11}\left({M_{\rm WD}\over 0.5{\rm\ M_\odot}}\right) {\rm\ cm} ,
\end{equation}
where $r_{i}$ is the distance of the innermost \ion{O}{VI} emission nebula from the white dwarf and $M_{\rm WD}$ is
the mass of the white dwarf.

\begin{figure*}
\centering
\includegraphics[scale=.40]{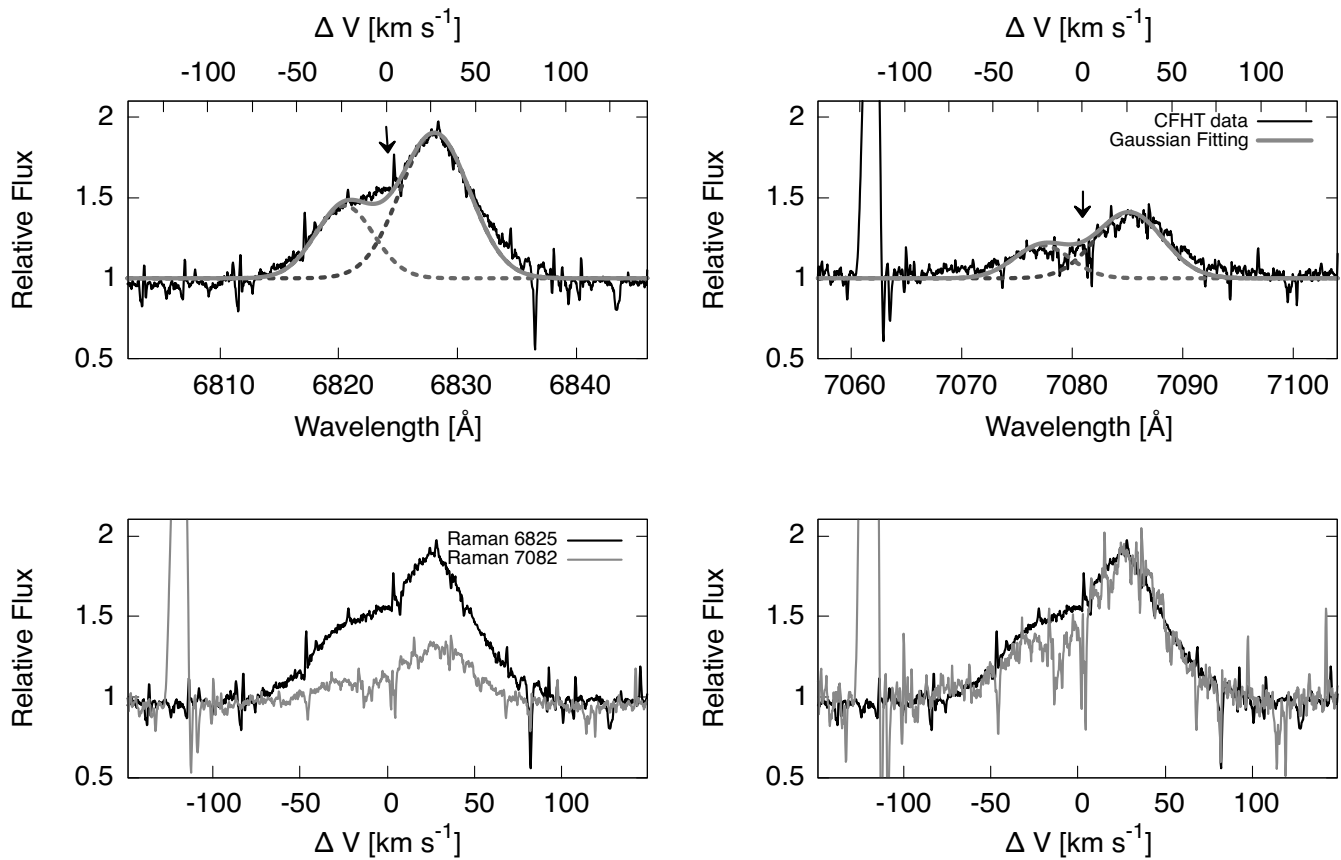}
\caption{ 
The upper panels show the double Gaussian fit to the {\it CFHT} spectrum.
The upper horizontal axis shows the Doppler factor in units of ${\rm km\ s^{-1}}$
of parent \ion{O}{VI} lines.
In the bottom left panel, the two Raman features are overplotted on the Doppler factor space.
In the bottom right panel, the flux of Raman 7082 feature is multiplied by 2.5 in order to make the two red peaks coincide.
The two profiles are overall similar, except possibly in the velocity range $-25{\rm\ km\ s^{-1}}$ to 0 where the Raman 7082 flux is relatively 
weaker than the Raman 6825 counterpart.
}
\label{overplot}
\end{figure*}

In Fig.~\ref{overplot}, we overplot the two Raman spectral features in the Doppler factor space in order 
to make a comparison of the two profiles. The two upper panels show our double Gaussian fit to the profiles,
where the sum of two Gaussian functions are represented by gray lines.
Taking the emission line \ion{He}{i}$\lambda$7065 with the atomic line center at  7065.190 \AA\ 
as velocity reference, 
we determine the centers 
of the two Raman \ion{O}{VI} features, which are shown by a vertical arrow in the top two panels of Fig.~\ref{overplot}.
The upper horizontal axis shows the Doppler factors computed in the parent far UV \ion{O}{VI} lines.
Both Raman 6825 and 7082 features show a main peak shifted redward from the profile center by a Doppler factor
of $\sim 25{\rm\ km\ s^{-1}}$.

The double Gaussian decomposition is made first for the Raman 6825 feature. We find that
the center separation between the two Gaussian components is measured to 
be $\sim 57{\rm\ km\ s^{-1}}$. The full widths at half maximum of the blue and red Gaussian functions
are $\Delta\lambda = 8.15$ \AA\ which corresponds to the \ion{O}{VI} velocity width of $54{\rm\ km\ s^{-1}}$.
The red Gaussian is 1.19 times stronger than the blue counterpart. We divide the two Gaussian functions by 2.5 
and apply to the Raman 7082 feature.

If the double component profiles are attributed to the accretion disk around the white dwarf,
the outer radius of the accretion disk is roughly $r_{o}\sim 0.3{\rm\ au}$  with the
adopted white dwarf mass of $M_{\rm WD} = 0.5 {\rm\ M_\odot}$ \citep{mikolajewska95}.
It is also noted that the location of Doppler factor zero
appears to be shifted blueward with respect to the center of zero intensities for both the Raman features,
implying an additional contribution to the red part.

 For a clear comparison,
we multiply the flux of Raman 7082 feature by a factor 2.5 in the bottom right panel.
It is found  that the two profiles coincide within the uncertainty. Whereas the red parts of the two profiles 
match excellently, the Raman 7082  appears slightly weaker than the 6825 counterpart 
in the velocity range $-25{\rm\ km\ s^{-1}}<\Delta V <0{\rm\ km\ s^{-1}}$. 
the absorption features near $\Delta V\sim -20{\rm\ km\ s^{-1}}$ and insufficient data quality
prevent a definite conclusion .

\section{Formation of Raman Scattered O~VI at 6825 \AA\ and 7082 \AA}

\subsection{Ionization Front}

The slow stellar wind region in the vicinity of the giant is illuminated and 
photoionized by the far UV radiation from the hot white dwarf. The  ionization front 
is formed where the photoionization rate balances the recombination rate.
The former is closely associated with the mass accretion rate onto the white dwarf component 
whereas the latter is mainly determined by the mass loss rate $\dot M$ of the giant.
\cite{seaquist84} introduced an ionization parameter $X$ to characterize the ionization front, which is
\begin{equation}
X={4\pi\mu^{2}m_{\rm p}^{2}\over\alpha}aL_*(v_\infty/{\dot M})^{2}.
\end{equation}
Here, $L_*, \mu, \alpha, m_{\rm p}$ and $a$ are the hydrogen ionizing luminosity,  the mean molecular weight,  
the recombination coefficient, the proton mass and the binary separation, respectively.

In this work,  we adopt a simple spherically symmetric stellar wind model so called a $\beta$-law given by
\begin{equation}
{\bf v}(r) = v_\infty \left(
{1-{R_*\over r}}
\right)^\beta,
\end{equation}
where $v_\infty$ is the wind terminal speed, $R_*$ is the distance of the wind launching place.
In this work, we set $R_*=35 R_\odot$ which \cite{hric14} proposed as the radius of the giant component
of AG~Dra.

We also choose $\beta=1$ for simplicity. The density $n(r)$ distribution corresponding to the  $\beta$-law is given by
\begin{equation}
n(r) = n_* \left({R_*\over r}\right)^{-2}
\left(
{1-{R_*\over r}}
\right)^{-1}.
\end{equation}
Here, the characteristic number density is  
\begin{equation}
n_*= {\dot M\over
4\pi \mu m_{\rm p} R_*^2 v_\infty} =  1.5\times 10^{10} 
\left({\dot M_{-7}\over v_{20} R_{35}^2}\right)
 {\rm\ cm^{-3}},
\end{equation}
where we set $\mu=1.4$ and the scaled parameters  $\dot M_{-7}=\dot M/(10^{-7}{\rm\ M_\odot\ yr^{-1}})$, 
$v_{20}=v_\infty/(20{\rm\ km\ s^{-1}})$ and $R_{35}=R_*/(35R_\odot)$.

\begin{figure}
\centering
\includegraphics[scale=.3]{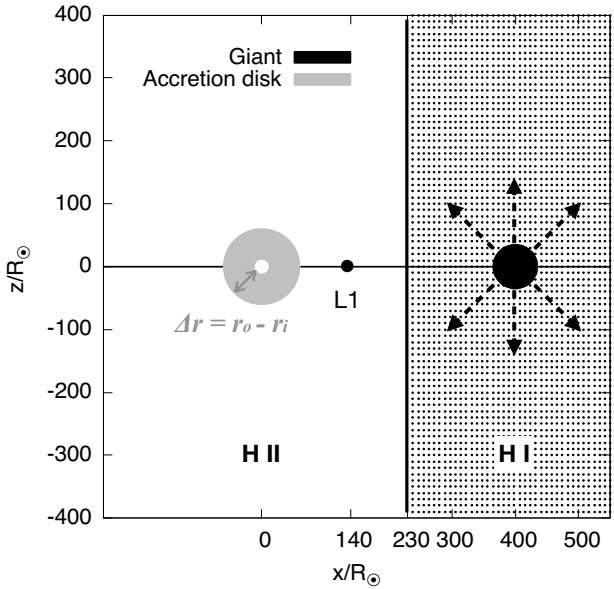}
\caption{A schematic illustration of the photoionization model of AG Dra
adopted in this work. We take the line connecting the white dwarf and the giant as the $x$-axis
and locate the white dwarf at the center of the coordinate system. The binary orbital plane coincides with
the $x-z$ plane.
The inner Lagrange point is located at 140 $R_{\odot}$ from the white dwarf, allowing an accretion
disk with a Keplerian speed $\sim 30 {\rm\ km\ s^{-1}}$ to be accommodated inside the Roche lobe of the white dwarf. 
The \ion{O}{VI} emission region is identified with the annular region with the inner and outer radii $r_{\rm i}$
and $r_{\rm o}$, respectively. The radial dotted arrows represent the slow stellar wind from the giant component.
}
\label{ionfront}
\end{figure}

In this stellar wind model,  neutral hydrogen column density
$N_{HI}$ diverges along lines of sight from the white dwarf toward the giant.  
On the other hand,  if the mass loss rate is moderate, complete photoionization results along lines of sight
with large impact factors from the giant. Therefore, one finds that the ionization front is well approximated by a hyperboloid
when the accretion luminosity is sufficient to fully ionize except the neighboring region of the giant component \citep[e.g.,][]{lee07, sekeras15}.
In this work, we consider mainly the case where the half-opening angle $\theta_o=90^\circ$ of the ionization front with respect to the white dwarf.
 Fig.~\ref{ionfront} shows a schematic illustration adopted in this work. 
 
 Taking the mass $M_{\rm G} = 1.5{\rm\ M_\odot}$ and radius $R_{\rm G}=35{\rm\ R_\odot}$, we show the position and size of the giant component to the scale
 in Fig.~\ref{ionfront} \citep{skopal05, hric14}.
In Table~\ref{mod_tab}, we summarize the model parameters of AG~Dra adopted in this work.
The inner Lagrange point is located at 140 $R_{\odot}$ from the white dwarf, allowing an accretion
disk with a Keplerian speed $\sim 30 {\rm\ km\ s^{-1}}$ to be accommodated inside the Roche lobe of the white dwarf.
The $x-$axis is chosen to coincide with the binary axis. The giant and the white dwarf are located at
$x=400\ R_\odot$ and at the center, respectively.
We put an additional \ion{H}{i} region along $x-$axis to mimic the bipolar structure of AG~Dra. 

\begin{table*}
\caption{
          Model parameters of AG Dra adopted in this work.
         }
\label{model_param}
\begin{tabular}{ccccc}
\hline
\hline
       Parameter & Value & Description &  Reference& \\
\hline
\hline
        $M_{\rm G}$   & 1.5 M$_{\odot}$& Mass of giant& \cite{skopal05}\\
        $R_{\rm G}$   & 35 R$_{\odot} $& Radius of giant& \cite{hric14}\\
        $M_{\rm WD}$& 0.5 M$_{\odot}$& Mass of white dwarf & \cite{mikolajewska95}\\
        a       & 400 R$_{\odot}$& Binary separation & Derived from \cite{fekel00}\\
        $L_1$   & 140 R$_{\odot}$& Inner Lagrange point & \\
	v$_\infty$ & 10, 15, 20 ${\rm km~s^{-1}}$ & Wind terminal velocity & \\
	$\dot M$ & 10$^{-5}$ - $10^{-7} $ ${\rm M_\odot~yr^{-1}}$ & Mass loss rate& \\
\hline

\end{tabular}
\label{mod_tab}
\end{table*}


\subsection{Monte Carlo Approach}

In this subsection, we briefly describe our Monte Carlo code to simulate Raman \ion{O}{VI} profiles.
The Monte Carlo simulation starts with generation of an \ion{O}{VI} line photon in the emission region around the white dwarf.
In view of the FWZI and the velocity separation between the two Gaussian components of the Raman features, 
we identify the \ion{O}{VI} emission region as 
the part of a Keplerian disk around the white dwarf with the inner and outer radii denoted by $r_i$ and $r_o$,
where these two parameters are described in Subsection 2.2.
 Using a uniform random number $u_r$ between 0 and 1, an O VI line photon is generated initially at a position 
with radial coordinate $r_1$ by the prescription
\begin{equation}
r_1 = [r_i^2+(r_o^2-r_i^2)u_{r}]^{1/2}.
\end{equation}
As a check of the code, we consider a point-like and a ring-like \ion{O}{VI} emission source by setting $r_i=r_o=0$
and $r_i=r_o$ to a positive constant, respectively. For these cases, the \ion{O}{VI} emissivity is assumed
to be isotropic. However, the observed red enhanced profiles require anisotropic \ion{O}{VI} emissivity,
which is described in subsection 5,1.

The initial unit wavevector ${\bf\hat k_i}$ is chosen isotropically. 
To the initial photon at ${\bf r}_1$ with ${\bf\hat k_i}$, we
assign a Doppler factor $DF$ given by
\begin{equation}
DF \equiv { {\bf\hat k_i}\cdot {\bf V}_{\rm Kep} \over c} ,
\end{equation}
where ${\bf V}_{\rm Kep}$ is the Keplerian velocity vector associated with the \ion{O}{VI} line emitter.
In the case of a point-like emission region, ${\rm V}_{\rm Kep}$ is set to be zero.

We trace photons entering the \ion{H}{i} region,
where Rayleigh or Raman scattering may take place. According to the total scattering cross section 
$\sigma_{\rm tot}=\sigma_{\rm Ray}+\sigma_{\rm Ram}$, a next scattering site
is determined in a probabilistic way. The scattering type is determined according to the branching ratios $b_{\rm Ram}=
\sigma_{\rm Ram}/\sigma_{\rm tot}$  and $b_{\rm Ray}=1-b_{\rm Ram}$.
If the scattering is Raman, then the Raman photon is presumed to escape from the region to reach the observer. 
Otherwise, we generate a new unit wavevector ${\bf\hat k_f}$ along which we find a new scattering site 
corresponding to a scattering optical depth $\tau$.
A new Doppler factor is assigned to the new ${\bf\hat k_f}$ and the velocity of the scatterer, which is given by
\begin{equation}
DF' = { {\bf\hat k_f}\cdot {\bf v}({\bf r_1}) \over c}.
\end{equation}
It may occur that no such scattering site is found when $\tau$ exceeds the optical depth corresponding to
the infinity, in which case  the \ion{O}{VI} line photon escapes through Rayleigh scattering. 
A scattering loop 
ends either with Raman escape or with Rayleigh escape. For a Raman-escaping photon, the final wavelength $\lambda_{\rm f}$
is given by
\begin{equation}
{\lambda_{\rm f}-\lambda_{o}\over\lambda_{o}}= \left({\lambda_{\rm o}\over\lambda_{\rm i}} \right)DF+DF'.
\end{equation}

The probabilistic determination of a next scattering site ${\bf r}_2$
from a starting position ${\bf r}_1$ and a unit wavevector ${\bf\hat k}$ involves the inversion of
the physical distance $s$ in terms of a total scattering optical depth $\tau$, which we describe now.

The characteristic total scattering optical depth $\tau_*$ is defined as
\begin{equation}
\tau_* (\lambda)=n_*  R_* \sigma_{\rm tot}(\lambda).
\end{equation}
The numerical values for \ion{O}{VI} are
$\tau_*(1032)=1.02$ and $\tau_*(1038)=0.22$.
Given a unit wavevector ${\bf\hat k}$, the scattering optical depth between two positions ${\bf r}_1$ and ${\bf r}_2$ is given 
by the line integral
\begin{equation}
\tau_{12}=\int_{{\bf r}_1}^{{\bf r}_2} n({\bf r})\sigma_{\rm tot}(\lambda) ds,
\end{equation}
where the parameter $s$ measures the physical distance from ${\bf r}_1$ to ${\bf r}_2$, that is, ${\bf r_2}={\bf r}_1+s{\bf\hat k}$ . 
We introduce the impact parameter $b$ that measures the perpendicular distance from the giant to the photon path.
In terms of $\tau_*$, the integral of interest can be written as
\begin{equation}
\tau(\rho)=\tau_*\int_{\rho}^{\infty}{d\rho\over (\rho-1)\sqrt{\rho^2-\tilde b^2}},
\end{equation}
where the dimensionless parameters are $\rho=r/R_*$ and $\tilde b=b/R_*$ \citep{lee97}. 

The inverse relation for the physical distance corresponding to a scattering optical depth $\tau$ is
\begin{equation}
\rho = \tilde b\cosh\theta(\tau) -1.
\end{equation}
Here, the parameter $\theta(\tau)$ is
\begin{eqnarray}
\theta(\tau) &=& 2\tanh^{-1}\left[
\sqrt{1-\tilde b\over 1+\tilde b} \coth\left(
\coth^{-1}\sqrt{1+\tilde b\over 1-\tilde b} 
\right. \right.
\nonumber \\
&& \left. \left.
+{\sqrt{1-\tilde b^2}\over2}{\tau\over\tau_*}
\right) \right],
\end{eqnarray}
when $\tilde b<1$. For $\tilde b>1$, the corresponding formula is
\begin{eqnarray}
\theta(\tau) &=& 2\tanh^{-1}\left[
\sqrt{\tilde b-1\over \tilde b+1} \coth\left(
\cot^{-1}\sqrt{\tilde b+1\over \tilde b-1} 
\right.  \right.
\nonumber \\
&& \left.  \left.
+{\sqrt{\tilde b^2-1}\over2}{\tau\over\tau_*}
\right)  \right].
\end{eqnarray}

 The red enhanced profiles shown in Fig.~\ref{btsettl} imply that the accretion disk
is asymmetric in the azimuthal direction. Hydrodynamic simulations show that the accretion flow tends to converge
on the entrance side as the wind material from the giant is captured by the white dwarf. However, on the opposite side
the flow tends to diverge so that the \ion{O}{VI} density diminishes leading to weaker \ion{O}{vi} emission
than the entrance side \citep[e.g.][]{devalborro09}.

\section{Monte Carlo Radiative Transfer and Basic Results}

\subsection{Raman Conversion Efficiency}

\begin{figure}
\centering
\includegraphics[scale=.23]{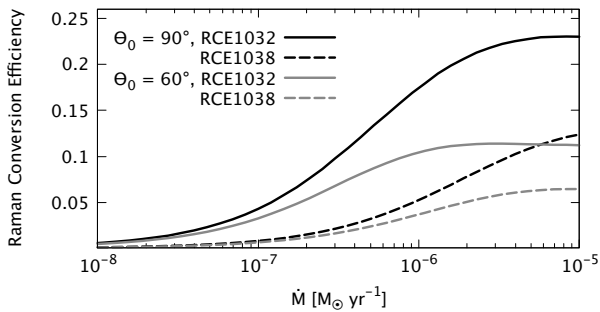}
\caption{Raman conversion efficiencies for \ion{O}{VI}$\lambda\lambda$1032 and 1038 and the flux ratio $F(6825)/F(7082)$
from a point-like and monochromatic \ion{O}{VI} source for various values of the mass loss rate $\dot M$ of the giant.
We set the  binary separation $a=400 {\rm R_\odot}$ and the half-opening angle $\theta_o=90^\circ$ of the ionization front. 
Here, we inject the same number of 1032 and 1038 line photons to compute the flux
ratio $F(6825)/F(7082)$.
}
\label{conversion}
\end{figure}

In this subsection, we compute Raman conversion efficiency (RCE)
for \ion{O}{VI}$\lambda\lambda$ 1032 and 1038 as a function of
the mass loss rate of the giant and the half-opening angle of the ionization front.  
Here, we define RCE as the number ratio of far UV incident photons generated in the \ion{O}{VI} emission region
and Raman scattered photons formed in the \ion{H}{i} region.
With the availability of far UV spectroscopy, Raman conversion
efficiency is a good measure of the mass loss rate of the giant component.

\cite{birriel00} proposed RCE of 15 percent for \ion{O}{VI}$\lambda$1032 from their near simultaneous 
observations of far UV using the {\it Hopkins Ultraviolet Telescope} and optical spectroscopy. 
\cite{schmid99} used the data obtained with the {\it Orbiting \& Retrievable Far and Extreme Ultraviolet Spectrometer} 
to deduce much higher efficiency of 50 percent with quoted error of $\sim$ 20 percent. \cite{birriel00} 
discussed possible origins for the discrepancy including the simultaneity of far UV and optical spectroscopy
and interstellar extinction.

\cite{schmid96} carried out Monte Carlo simulations to study basic properties of Raman scattering of \ion{O}{VI} with atomic
hydrogen and  estimate the RCE of \ion{O}{VI} doublets in symbiotic stars. 
\cite{ylee16} investigated the Raman conversion efficiencies of 
\ion{O}{VI}$\lambda\lambda$ 1032 and 1038 for a few simple
scattering geometries to find that the flux ratio $F(6825)/F(7082)$ may vary in the range from 1 to 6.
 
In Fig.~\ref{conversion}, we show the RCE for \ion{O}{VI}$\lambda\lambda$ 1032, 1038 
and the flux ratio $F(6825)/F(7082)$.
The binary separation is set to $a=400{\rm\ R_\odot}$ and the mass loss rate of the giant
is taken to be in the range $10^{-8}-10^{-6}{\rm\ M_\odot\ yr^{-1}}$. 
We present our results for the two values $\theta_o=60^\circ$ and $90^\circ$ of  the half-opening angle of the ionization front.

The overall RCE increases monotonically as $\dot M$ increases.
The maximum Raman conversion efficiency of \ion{O}{VI}$\lambda$1032 is found to be 0.23 and 0.11 for $\theta_o=90^\circ$ 
and $\theta_o=60^\circ$, respectively.
For \ion{O}{VI}$\lambda$1038, smaller values of 0.13 and 0.07 are obtained. When the mass loss rate is high, RCE becomes very sensitive
to the half-opening angle. However, when the mass loss rate is low $\le 10^{-7}{\rm\ M_\odot\ yr^{-1}}$, RCE is mainly determined by
$\dot M$ and almost independent of $\theta_o$. In the case of low mass loss rates, most of Raman photons are formed near the giant,
where the optical depth diverges irrespective of $\dot M$ in our model. It is interesting to note that the RCE of 15 percent of \ion{O}{VI}$\lambda$1032
proposed by \cite{birriel00} is consistent with a mass loss rate $\dot M\simeq 4\times10^{-7}{\rm\ M_\odot \ yr^{-1}}$ with $\theta_o=90^\circ$.

\subsection{Pure $\beta$-Law for the Giant Stellar Wind}

\begin{figure*}
\centering
\includegraphics[scale=.4]{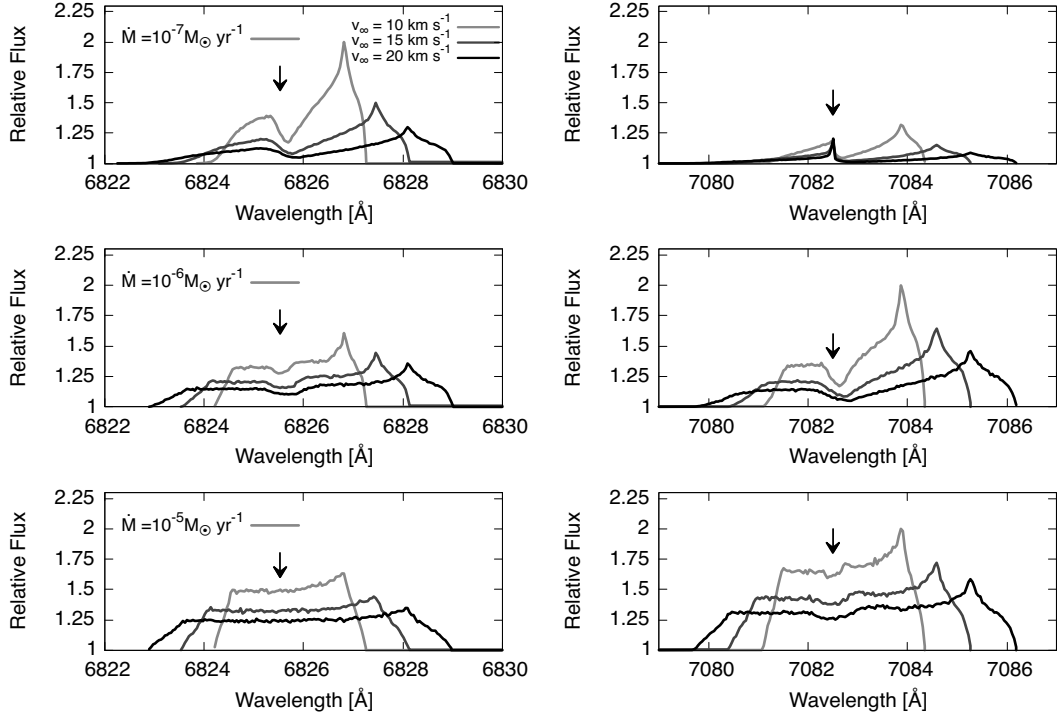}
\caption{Raman scattered 6825 and 7082 features from a point-like monochromatic \ion{O}{VI} source located near the
white dwarf for various values of the mass loss rate $\dot M$ ranging from $10^{-7}{\rm\ M_\odot\ yr^{-1}}$
to $10^{-5}{\rm\ M_\odot\ yr^{-1}}$. The stellar wind from the giant is assumed to follow
the $\beta$ law with $\beta=1$ and the terminal giant wind speed $v_\infty$, which is varied from $10{\rm\ km\ s^{-1}}$
to $20{\rm\ km\ s^{-1}}$. 
}
\label{monochromat}
\end{figure*}

\begin{figure*}
\centering
\includegraphics[scale=.3]{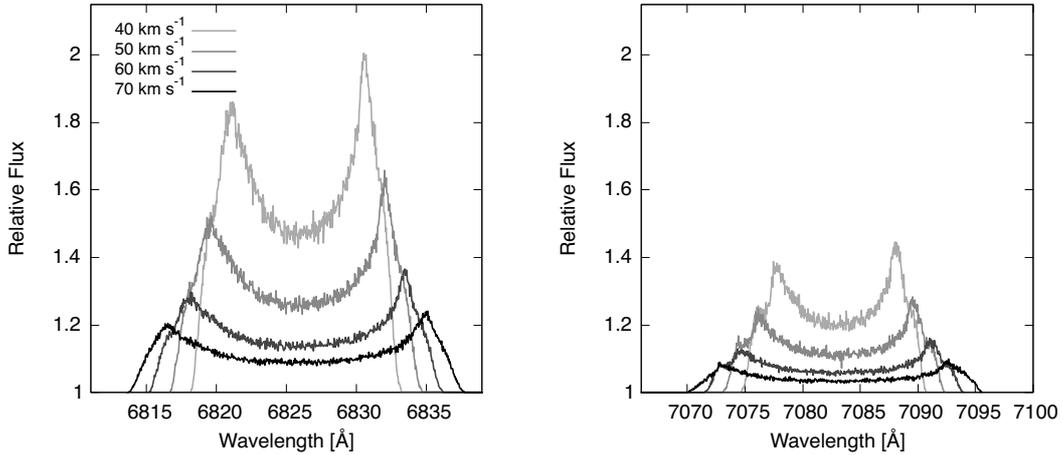}
\caption{Monte Carlo simulated Raman \ion{O}{VI} profiles at 6825 \AA\ (left panels), 7082 \AA\ (right panels) from
an \ion{O}{VI} emission source that forms a uniform ring around the white dwarf. The parameters used are
the giant's mass loss rate $\dot M=4 \times 10^{-7}{\rm M_\odot \ yr^{-1}}$,
the half-opening angle of the ionization front $\theta_o=90^\circ $ and the binary separation of $400{\rm\ R_\odot}$.
The terminal velocity of the giant wind is set to $v_\infty=10{\rm\ km\ s^{-1}}$. 
Four values 40, 50, 60 and $70 {\rm\ km\ s^{-1}}$ of ${\bf V}_{\rm Kep}$ are chosen, for which
the results are shown with different line types.
}
\label{ring}
\end{figure*}

In Fig.~\ref{monochromat}, we show the profiles of Raman scattered \ion{O}{VI} features formed in a stellar wind with $\beta=1$
for various values of the mass loss rate $\dot M$ and the giant wind terminal speed $v_\infty$. 
In this figure, the \ion{O}{VI} emission region is taken to be point-like and monochromatic.
The vertical arrow indicates the atomic Raman line center, that is the wavelength a line center photon 
would acquire if it is Raman scattered photons in a static \ion{H}{i} region.

We consider three values of the terminal wind velocity $v_\infty=10, 15$ and $20{\rm\ km\ s^{-1}}$.
As the wind terminal speed is increased, the profile width increases. At a specified neutral region around the giant,
the density is proportional to $\dot M$ and inversely proportional to $v_\infty$ so that the Raman flux
decreases as $v_\infty$ increases given mass loss rate.

We note that the profile peak is found significantly redward of the Raman line center. This is also found in the work of \cite{lee97},
who also presented the contour of the Raman scattering optical depth in their Fig.~2. The contour of the Raman scattering optical depth of unity
takes approximately a hyperboloidal shape, which shows that \ion{O}{VI} line photons are scattered more efficiently in the stellar wind region
receding from the white dwarf. The red peaks are found at $\Delta\lambda=+1.3, 1.9$ and 2.6 \AA\ from the atomic line center
for Raman scattered 6825 feature for $v_\infty = 10, 15$ and $20{\rm\ km\ s^{-1}}$, respectively.
These red peaks turn out to correspond to $v=0.75 v_\infty$.

In addition, there appears a blue shoulder in the cases where $\dot M\le 1\times 10^{-6}{\rm\ M_\odot\
yr^{-1}}$. This is due to contribution
from the part of the giant stellar wind facing the white dwarf, where Raman scattering optical depth increases without limit for 
lines of sight hitting the giant surface.  For $\dot M>10^{-6}{\rm\ M_\odot}$, Raman scattering occurs quite far from the giant surface, where
the Raman scattering optical depth is moderate. Therefore, the neutral region near the giant surface facing the white dwarf contributes little,
and no shoulder features are found in the bottom panels of Fig.~\ref{monochromat}.

Fig.~\ref{ring} shows the profiles and relative fluxes of Raman 6825 and 7082 features where the \ion{O}{VI} emission
region is assumed to form a uniform ring around the white dwarf.
We consider four values of the Keplerian rotation velocity ${\bf V}_{\rm Kep}$ of the \ion{O}{VI} emission region, which are 40, 50 ,60 and $70{\rm\ km\ s^{-1}}$. 
In this figure, we fix the mass loss rate of the giant $\dot M= 4 \times 10^{-7}\ {\rm M_\odot \ yr^{-1}}$,
the half-opening angle of the ionization front $\theta_o=90^\circ $ and the binary separation of $400{\rm\ R_\odot}$.
The terminal velocity of the giant wind is set to $v_\infty=10{\rm\ km\ s^{-1}}$.

The two main peaks correspond to the rotation speed of the \ion{O}{VI} emission ring. 
As in the case of Fig.~\ref{monochromat},
the red peaks are slightly stronger than the blue counterparts. A close look reveals that the full width at zero intensity
is slightly larger than the peak separation due to convolving effect of the neutral wind around the giant.
We notice that there appears slight excess in the relative flux near the blue and red ends, which is also attributed 
to the effect of convolution.
It should be noted that the overall profile is mainly determined by the kinematics of the emission region if the giant wind
speed is small compared to the profile width. The conspicuous red enhancement apparent in the observational data
strongly implies that the \ion{O}{VI} emission region is significantly azimuthally asymmetric.

\section{ Raman O~VI Profiles and Comparisons with Observation}

\subsection{Raman Flux Ratio and the Mass Loss Rate}
\begin{figure*}
\centering
\includegraphics[scale=.45]{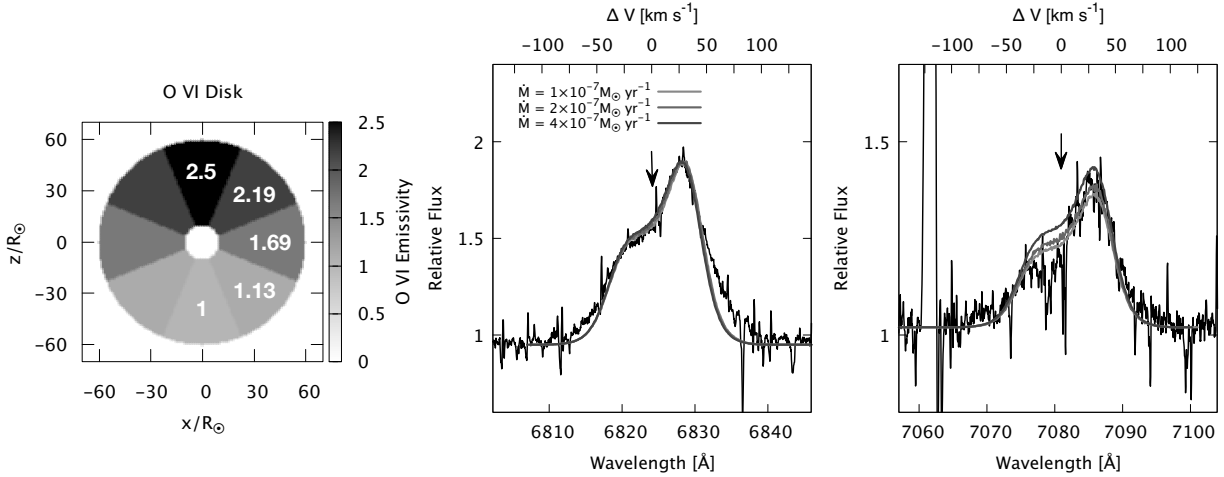}
\caption{Relative flux and profiles of Raman scattered 6825 and 7082 features obtained with our Monte Carlo calculations
for three values of mass loss rate $\dot M$ of the giant component. The center and right panels show Raman 6825 and 7082 profiles,
respectively. \ion{O}{VI} emission originates from an asymmetric accretion disk, where the emissivity is given 
as a function of only azimuthal angle $\phi$. The left panel shows the emissivity by gray scale. 
}
\label{mdotdep}
\end{figure*}

In this subsection, we consider an annular \ion{O}{VI} emission region depicted schematically in Fig.\ref{ionfront}.
For simplicity, we assume that
the \ion{O}{VI} local emissivity $\epsilon({\bf r})$ is a function of only azimuthal angle $\phi$
measured from the line toward the giant component.
We further assume that the flux ratio $F(1032)/F(1038)$ is constant throughout the annular \ion{O}{VI}
region. 

Dividing the entire azimuthal angle domain into
8 equal intervals, each being centered at $\phi_j=j \pi/4$, $(j=0,1,2,\dots, 7),$ we try piecewise step functions 
taking a constant value in each interval. That is,
\begin{equation}
\epsilon({\bf r}) = \epsilon(\phi)=A_j, \quad \phi\in (\phi_j-\pi/8,\ \phi_j+\pi/8).
\end{equation}

Once we find a set of parameters $A_j$ fitting the observed Raman 6825 data, the same set of parameters  are used
to fit the Raman 7082 feature. In the center panel of Fig.~\ref{mdotdep}, we show our best
fitting profiles for Raman 6825 for three different values of $\dot M =1\times 10^{-7},\ 2\times 10^{-7}$ and $4\times10^{-7}
{\rm\ M_\odot\ yr^{-1}}$. The criterion for best fit adopted in this subsection is the least chi-square 
in the wavelength range between 6815 \AA\ and 6830 \AA\ that encompasses the blue shoulder and the red main peak. 

The best fitting values of $A_j$ are listed in Table~\ref{emi_tab} and we also illustrate $\epsilon(\phi)$ 
in the left panel of Fig.~\ref{mdotdep}.
In the figure, it is noted that
the \ion{O}{VI} emissivity peaks at around $\phi=90^\circ$, where it is 2.5 times larger than at $\phi=270^\circ$.

 \begin{table*}
  \caption{ \ion{O}{VI} local emissivity $\epsilon({\bf r})$ as a function of the azimuthal angle in the accretion disk. }
  \label{loc_emi}
  \begin{tabular}{| ccccccccc | }
  \hline
  $\phi_j$ & $0$ & $ \pi/4$ & $\pi/2$ & $3\pi/ 4
  $ & $\pi$& $5\pi/4$& $3\pi /2$ & $7\pi/4$ \\ \hline

  $A_j$ & 1.69 & 2.19 & 2.5 & 2.19 & 1.69 & 1.13 & 1 & 1.13 \\ \hline
 \end{tabular}
\label{emi_tab}
  \end{table*}

In the right panel of Fig.~\ref{mdotdep}, we present three profiles simulated for Raman \ion{O}{VI}$\lambda$7082.  
Our {\it CFHT} spectra are overplotted to illustrate the quality of the fit in the center and right panels.
The main effect of the mass loss rate $\dot M$ is found in the flux ratio of $F(6825)/F(7082)$, where the flux ratio 
tends to decrease as $\dot M$ increases. 

The simulated profiles of Raman 6825 features fail to fit the red part, which is also clearly linked 
to the blueward shift of the atomic line center. 
It appears that the observed Raman 7082 feature is best fit with $\dot M=2\times10^{-7}{\rm\ M_\odot\ yr^{-1}}$.
However, the difference between simulated profiles for various values of $\dot M$ is not large compared to the 
uncertainty of the observational data around 7082 \AA, from which it is very difficult to single out a definite value 
of $\dot M$. Further studies including polarimetric observations and spectroscopic monitoring will provide 
additional constraints on the mass loss rate and mass transfer processes.

\subsection{Bipolar Structure}

\begin{figure*}
\centering
\includegraphics[scale=.26]{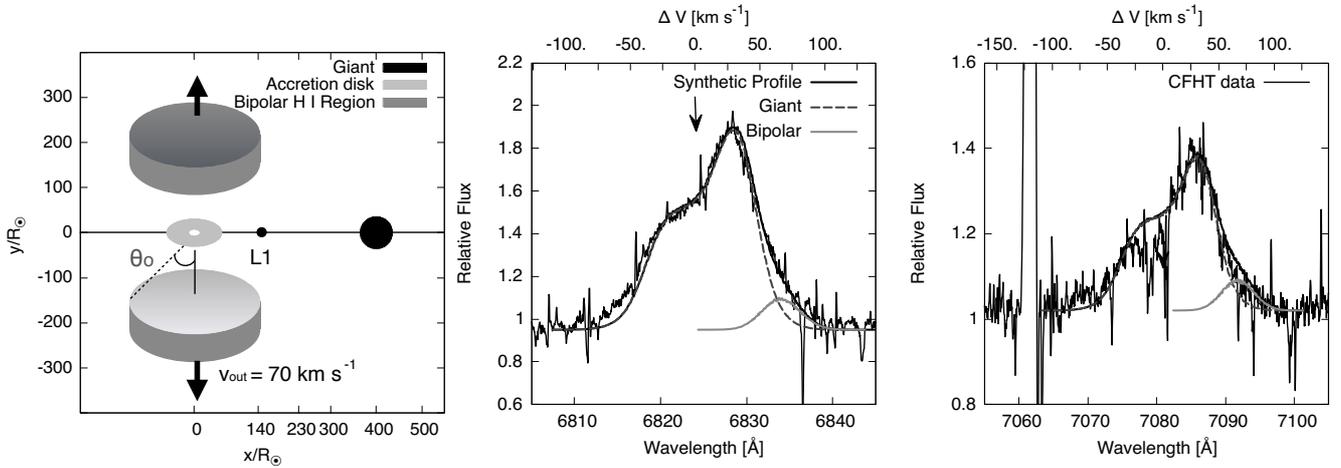}
\caption{Profiles shown by a thick solid line are obtained from a simulation 
with an additional contribution from an \ion{H}{i} component moving away from the binary
system in the directions perpendicular to the orbital plane. The bipolar \ion{H}{i} components are receding 
with a speed $70{\rm\ km\ s^{-1}}$. The mass loss rate $M_{\odot}=2\times 10^{-7}{\rm\ M_\odot\
yr^{-1}}$. For comparison, the profiles shown by a dashed
line are obtained from the accretion model without bipolar \ion{H}{I} regions  illustrated in the bottom center and right panels of Fig.~\ref{mdotdep}.}
\label{bipolar}
\end{figure*}

The red excess apparent in the Raman \ion{O}{vi} profiles indicates the operation
of Raman scattering between \ion{O}{VI} and \ion{H}{i} components that are receding from each other. 
\cite{torbett87} presented
their radio 6 cm observation of AG~Dra to reveal a bipolar structure present in this system. It appears that
bipolar structures are commonly found in symbiotic stars, where collimated outflows are formed in association
with the mass accretion on to the white dwarf component \citep[e.g.,][]{angeloni11,skopal13}.
In this subsection, we further consider the presence of an \ion{H}{i} component moving away from the binary
system, from which an additional contribution is made to the red wing flux of Raman scattered \ion{O}{VI}.

The asymmetric wind accretion model provides a reasonable fit to the observational data with the exception
that the red wing parts are observed stronger than model fluxes.  In order to augment the deficit in the red wing 
parts for both simulated Raman features, it is necessary to introduce an additional
\ion{H}{i} region that is moving away from the two stars in the direction perpendicular to the orbital plane.

A significant fraction of planetary nebulae exhibit bipolar morphology, of which the physical origin is still
controversial \citep[e.g.,][]{sahai07}. Possible mechanisms include the binarity of the central star system
and magnetic fields. 
\cite{akras17} reported the detection of near IR emission from ${\rm H}_2$ in the bipolar planetary nebula K4-47, 
which delineates the low-ionization structures. They invoked the interaction of collimated flows or bullets that
are moving away from the central star along the bipolar directions. Interestingly enough, the receding velocity
of the low-ionization structures is typically of order $100{\rm\ km\ s^{-1}}$. 

In this subsection, we introduce bipolar \ion{H}{i} regions moving away in the directions perpendicular 
to the orbital planes.
The left panel of Fig.\ref{bipolar} shows a schematic illustration of the scattering geometry with an
addition of bipolar \ion{H}{I} regions.
The half-opening angle $\theta_o$ of the polar \ion{H}{i} region is chosen to be $45^{\circ}$
with a receding speed of $70{\rm\ km\ s^{-1}}$.
The \ion{H}{i} column density is set to $N_{HI}=10^{21}{\rm\ cm^{-2}}$.

In the center and right panels of Fig.~\ref{bipolar}, we show our simulated profiles with an additional 
contribution from a receding \ion{H}{i} region. 
The additional fluxes are shown in gray solid lines, constituting about 20 percent of the
total flux in both the Raman 6825 and 7082 features. 
One notable feature is that the additional Raman fluxes exhibit a single peak profile. This is due to the fact
that the velocity component parallel to the orbital plane does not contribute to the Doppler factor along the
perpendicular direction.
With this additional component, substantial improvement is achieved in the quality of fit in the red parts.

\subsection{Local Variation of O~VI Flux Ratio in the Accretion Flow}

\begin{figure*}
\centering
\includegraphics[scale=.45]{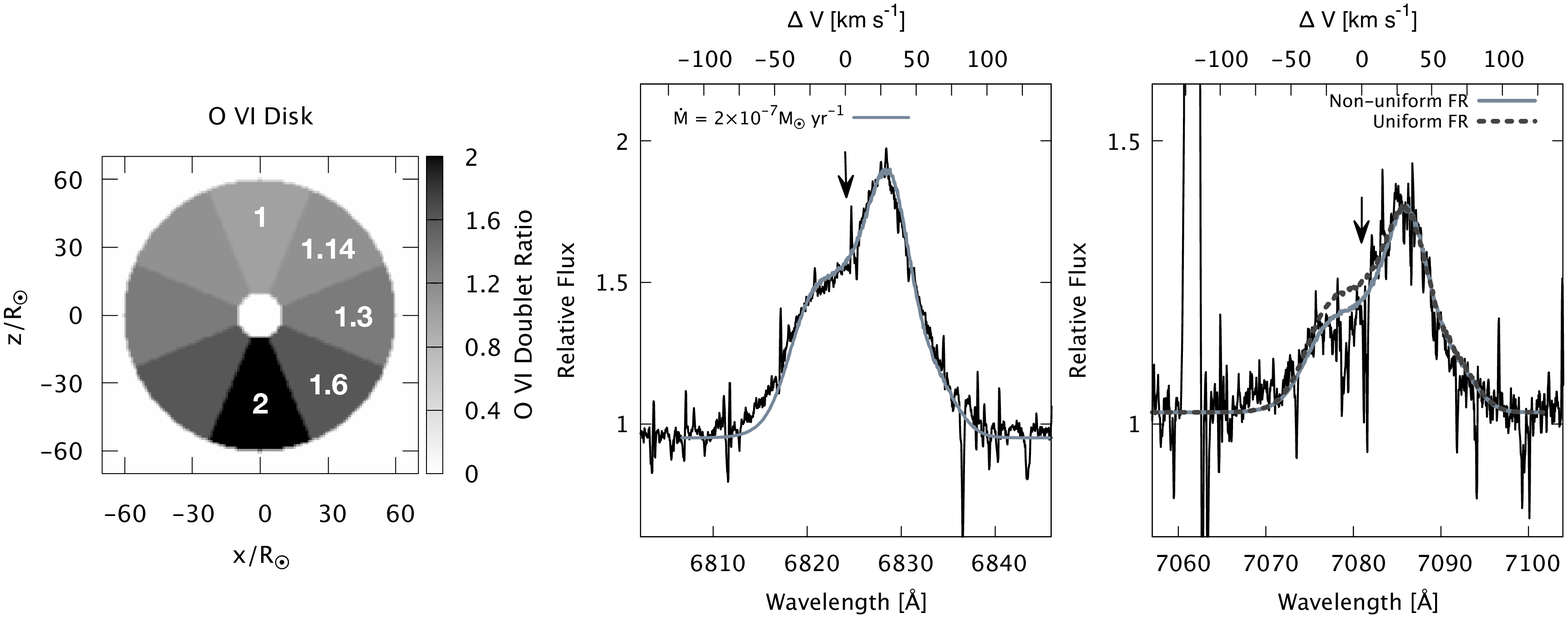}
\caption{Flux ratio $F(1032)/F(1038)$ as a function of $\phi$
in the \ion{O}{VI} emission region (left panel). The flux ratio $F(1032)/F(1038)\simeq 1$ in the red emission region and
$\simeq2$ in the blue emission region.
 The best fitting profiles are shown in the center and right panels in solid lines. The dashed lines show the profiles
shown in Fig.\ref{bipolar}, in which the flux ratio is uniform throughout the \ion{O}{VI} emission region. 
}
\label{fluxratio_vary}
\end{figure*}

In Subsection~2.2, we discussed the possibility that 
 the blue part of the Raman 7082 feature may be relatively more suppressed than in the Raman 6825 counterpart.
The disparity of the two Raman \ion{O}{VI} profiles in symbiotic stars including V1016~Cyg has been pointed out by a number of
researchers \citep[e.g.,][]{schmid99}. The disparate profiles are attributed to locally varying
flux ratios of $F(1032)/F(1038)$ in addition to the asymmetric
accretion flow around the hot white dwarf component \citep{heo15}. 

Resonance doublet lines including \ion{C}{IV}$\lambda\lambda$1548, 1551 and \ion{O}{VI}$\lambda\lambda$1032, 1038
arise from transitions between $S_{1/2}-P_{3/2, 1/2}$. The short wavelength component $S_{1/2}-P_{3/2}$
has twice larger statistical weight than the long wavelength component, leading to the flux ratio of 2 in an 
optically thin nebula.
Far UV spectroscopy of symbiotic stars and planetary nebulae reveals that resonance doublet lines
rarely show the flux ratio of 2. Instead their flux ratios are observed to
vary from one to two \citep{feibelman83}.  A similar result was reported by \cite{vassiliadis96}, who investigated
 a dozen planetary nebulae in the Magellanic Clouds using the {\it Hubble Space Telescope}.

In the accretion flow model shown in Fig.~\ref{ionfront}, we pointed out that the entering side is denser
than the opposite side. 
Nonuniform \ion{O}{VI} density distribution may lead to flux ratio $F(1032)/F(1038)$
that vary locally in the \ion{O}{VI} region. As an exemplary case, we assume that 
flux ratio $FR=F(1032)/F(1038)$ is given as a function of only azimuthal angle $\phi$.

As in the previous section for $\epsilon(\phi)$, we try piecewise step functions for $FR(\phi)$ given
as
\begin{equation}
FR(\phi) = B_j, \quad \phi\in (\phi_j-\pi/8,\ \phi_j+\pi/8),
\end{equation}
where $B_j\  (j=0,1,2,\dots, 7)$ are constants between 1 and 2.
In our model, $FR(\phi)$ is close to 2 in the blue region with $180^\circ <\phi < 360^\circ$
whereas it is near one in the red region with $0<\phi< 180^\circ$.


  \begin{table*}
  \caption{
  Flux ratio $FR(\phi)=F(1032)/F(1038)$ as a piecewise step function of azimuthal angle $\phi$ in the \ion{O}{VI}
 annular region.
  }
    \label{loc_fr}
    \begin{tabular}{| ccccccccc | }
    \hline
    $\phi_j$ & $0$ & $ \pi/4$ & $\pi/2$ & $3\pi/ 4
    $ & $\pi$& $5\pi/4$& $3\pi /2$ & $7\pi/4$ \\ \hline
    $B_j$  & 1.3 & 1.14 & 1 & 1.14 & 1.3 & 1.6 & 2 & 1.6 \\ \hline
   \end{tabular}
\label{fr_tab}
   \end{table*}

In the left panel of Fig.~\ref{fluxratio_vary}, we illustrate $FR(\phi)$ and the values of $B_j$ are listed
in Table~\ref{fr_tab}.
In the center and right panels, we show our Monte Carlo profiles, which are overplotted on the observational data.
Here, the dashed line represents the profiles shown in Fig.~\ref{bipolar}
for comparison, in which the flux ratio is uniform throughout the \ion{O}{VI} region. As is expected,
the resultant Raman 7082 profile shows significant deficit in the region
$-30{\rm\ km\ s^{-1}}\le \Delta V \le 0$ compared to the best fit profile obtained 
from the uniform flux ratio model, achieving enhancement in quality of profile fitting. 
Future high resolution spectroscopy with higher signal to noise ratio is expected to
resolve the issue of the profile disparity and the local variation of the flux ratio.

\section{Summary and Discussion}

In this work, we present our high resolution spectrum around Raman scattered \ion{O}{VI} bands 
of the yellow symbiotic star AG~Dra obtained with the {\it CFHT}. Both the Raman 6825 and 7082
features exhibit asymmetric double component profiles. We  find that
 in the reference frame of \ion{He}{i}$\lambda$7065, the atomic line centers of both Raman 6825 and 7082 features 
are shifted blueward, implying an additional contribution from components that recede from the two stars.

We have carried out profile analyses by performing Monte Carlo simulations.
The best fit model includes an accretion disk with a size $<0.3$ au around the white dwarf, a slow stellar
wind from the giant with a mass loss rate $\sim 2\times10^{-7}{\rm\ M_\odot}\ {\rm yr^{-1}}$
and neutral regions moving away from the binary orbital plane in the perpendicular directions with a speed $\sim
70{\rm\ km\ s^{-1}}$.

The observed line profiles of Raman \ion{O}{VI} 6825 and 7082 features exhibit overall redward shift with respect
to their expected atomic line centers. 
With this in mind we introduced bipolar neutral regions that recede from the binary system along the direction
perpendicular to the orbital plane. The bipolar neutral component provides an additional contribution to the red wing part
of Raman \ion{O}{VI} features, offering significant improvement in overall quality of profile fitting. 
This additional component is naturally inferred from radio observations
indicative of the bipolar morphology of AG~Dra \citep{torbett87}.

\cite{ylee16} pointed out that the Raman conversion efficiency is also a sensitive function of wavelength
so that the efficiency of a blue 1032 photon is slightly larger than that of a red 1032 photon.   
A similar effect is also expected of \ion{O}{VI}$\lambda$1038.
Our current Monte  Carlo code
adopts a set of fixed values for Rayleigh and Raman cross sections for \ion{O}{VI}$\lambda$1032 and another
set for \ion{O}{VI}$\lambda$1038, neglecting the slight difference of the cross section depending on the wavelength. 
A careful incorporation of this effect into the simulation may strengthen slightly
the blue parts of Raman \ion{O}{VI} features, leading to further slight distortion of the resultant profiles. 

It should be noted that a number of factors may affect Raman \ion{O}{VI} including the variation of $\dot M$
due to pulsational instability of the giant. Mass transfer may also be subject to instability associated with
the accretion disk. It is also quite likely that the part of the \ion{H}{I} region illuminated by \ion{O}{VI} line
radiation and visible to the observer will be dependent on the binary orbital phase. Our current Monte
Carlo code does not include these subtle points, which will be considered in the future works.

Raman-scattered features consist of purely scattered photons without any dilution from unpolarized incident radiation.
Due to this inelasity and the dipole nature of Raman scattering, the 6825 and 7082 features may develop strong 
linear polarization in the direction perpendicular to the scattering plane.
\cite{harries96} carried out spectropolarimetry of many symbiotic stars to show that Raman \ion{O}{VI} features 
are generally polarized with large degree of polarization.
They also showed that the position angle in the red wing parts
differs by an amount of $90^\circ$ from the main parts for both Raman \ion{O}{VI} features,
which is consistent with the bipolar structure inherent to symbiotic stars \citep[e.g.,][]{lee99}. 

According to \cite{schmid97},  AG~Dra exhibits rotation of polarization angle by an amount of 45 degrees 
in the Raman \ion{O}{VI} features. That is, the incident direction for red wing photons may differ by the same angle
to the line connecting the white dwarf and the giant. Our model strongly implies that the position angle differs 
by $90^\circ$. A more sophisticated scattering geometry including additional \ion{H}{I} regions
receding in oblique directions, which is deferred to future works.
Further high resolution spectropolarimetry will shed much light on the scattering geometry and mass loss process of AG~Dra.

\section*{Acknowledgements}
We are grateful to the anonymous referee for constructive comments that helped improve the paper.
This research was supported by the Korea Astronomy and Space Science Institute
under the R\&D program(Project No. 2015-1-320-18) supervised by the Ministry 
of Science, ICT and Future Planning. 
This work was supported by the National Research Foundation of Korea(NRF) 
grant funded by the Korea government(MSIT) (No. NRF-2018R1D1A1B07043944).
This work was also supported by K-GMT
Science Program (PID: 14BK002) funded through Korea GMT
Project operated by Korea Astronomy and Space Science
Institute.
RA acknowledges financial support from the DIDULS Regular PR17142 by Universidad de La Serena.







\bsp	
\label{lastpage}
\end{document}